\documentclass{emulateapj}

\newcommand{\kms}{km~s$^{-1}$}
\newcommand{\kmsy}{km~s$^{-1}$~yr$^{-1}$}
\newcommand{\obj}{SDSS~J1536+0441}
\newcommand{\bla}{BL09}
\newcommand{\blb}{LB09}
\newcommand{\hal}{H$\alpha$}
\newcommand{\hb}{H$\beta$}

\begin{document}

\shorttitle{An Unusual Double-Peaked Emitter}
\shortauthors{Chornock et al.}

\title{The Quasar SDSS J1536+0441: An Unusual Double-Peaked Emitter}

\author{R. Chornock\altaffilmark{1,2},
J. S. Bloom\altaffilmark{1},
S. B. Cenko\altaffilmark{1},
A. V. Filippenko\altaffilmark{1},
J. M. Silverman\altaffilmark{1},
M. D. Hicks\altaffilmark{3},
K. J. Lawrence\altaffilmark{3},
A. J. Mendez\altaffilmark{4},
M. Rafelski\altaffilmark{4}, and
A. M. Wolfe\altaffilmark{4}
}

\altaffiltext{1}{Department of Astronomy, University of California,
                 Berkeley, CA 94720-3411.}
\altaffiltext{2}{\texttt{chornock@astro.berkeley.edu} .}
\altaffiltext{3}{Jet Propulsion Laboratory, JPL MS 183-501, 4800 Oak
  Grove Drive, Pasadena, CA 91109.}
\altaffiltext{4}{Department of Physics, and Center for Astrophysics
  and Space Sciences, University of California, San Diego, 9500 Gilman
  Drive, La Jolla, CA 92093-0424.}

\begin{abstract}
The quasar SDSS J153636.22+044127.0, exhibiting peculiar broad
emission-line profiles with multiple components, was proposed as a
candidate sub-parsec binary 
supermassive black hole system. More recently, imaging revealed two
spatially distinct sources, leading some to suggest the system to be a
quasar pair separated by $\sim$5 kpc.  We present Palomar and Keck
optical spectra of this system from which we
identify a third velocity component to the emission lines.  We argue
that the system is more likely an unusual member of the class of
active galactic nuclei (AGNs) known as ``double-peaked emitters'' than a
sub-parsec black hole binary or quasar pair.  We find no significant
velocity evolution of the two main peaks over the course of 0.95 yr,
with a $3\sigma$ upper limit on any secular change of 70~\kmsy.  We
also find that the three velocity components of the emission lines are
spatially coincident to within 0$\farcs$015 along the slit, apparently
ruling out the double-quasar hypothesis.
\end{abstract}


\section{Introduction}

When galaxies merge, it is expected that their central supermassive
black holes will form a binary which will also eventually merge via
the emission of gravitational radiation \citep{bbr80}.  The process by
which the orbit of the two black holes shrinks to the point at which
gravitational radiation is effective in producing a merger within a
Hubble time is not currently understood (dubbed the ``final parsec
problem''; e.g., \citealt{Milo03}), and the observational
signatures of such systems have proved to be elusive. 

   Boroson \& Lauer (2009; hereafter \bla) recently identified a candidate
sub-parsec binary supermassive black hole system which could
potentially merge in less than a Hubble time.  They searched the Sloan
Digital Sky Survey (SDSS; \citealt{dr7}) quasar spectra and identified
the object SDSS J153636.33+044127.0 (hereafter \obj) as one of two
low-redshift ($z<0.7$) quasars exhibiting components having multiple 
redshifts. The key features of the system are two components of the
broad emission lines separated by $\sim3500$~\kms.  They interpreted
these to be the result of broad-line regions around a pair of black
holes with masses of 10$^{7.3}$ and 10$^{8.9}$ M$_{\odot}$, with the
relative velocity being a result of the orbital motion in a binary having
a sub-parsec semimajor axis with a period of $\sim$100 years.

The night after the \bla\ publication came to our attention, we obtained
Palomar spectra and identified a third velocity component
of the broad emission lines \citep{chornockATEL}.  This led us to 
conclude that \obj\ was instead an unusual member of
the class of AGNs known as ``double-peaked emitters'' (e.g.,
Halpern \& Filippenko 1988; Eracleous
\& Halpern 1994; Strateva et al. 2003), a
suggestion that was later made independently by \citet{gaskell09}.

Radio observations by \citet{wrobel09} led those authors to suggest a third
possibility: that of a quasar pair.  They identified two radio sources
separated by 0$\farcs$97 in Very Large Array (VLA) observations 
of \obj.  Subsequently,
\citet{decarliIAUC} found that the two radio sources were coincident
with a pair of $K$-band sources that they suggested were
members of a $\sim$5~kpc quasar pair.  As we completed this
work, a second preprint by Lauer \& Boroson (2009; hereafter \blb)
appeared.  They presented {\it Hubble Space Telescope (HST)} images
confirming the double nature of the optical source, but they also
argued that the optical data were inconsistent with a double-quasar
hypothesis. 

In \S 2 of this Letter, we present the Palomar spectra that led to our
initial identification of this source as a double-peaked emitter, as
well as additional high-quality Keck spectra.  We set upper
limits on the long-term evolution of velocity shifts of the broad
peaks and on any spatial offset of the emission regions
in \S3.  We then argue (\S 4) that the double-peaked emitter nature of 
this object provides a better explanation than the alternatives. 
\blb\ reached similar conclusions, so we
compare with their results where appropriate and highlight our
differences.

\section{Observations}

Our initial spectroscopy was performed using the Double Spectrograph
on the Palomar 5-m Hale telescope \citep{oke82}; two 600~s exposures
were obtained 
on 2009 March 7.52 UT.  The blue-side (3150--5700~\AA) observations we
present here had a spectral resolution of $\sim$6~\AA\ from the
combination of a 1$\farcs$5 slit and the 300/3990 grating.
In addition, we obtained a pair of 600~s observations
with the Echellette Spectrograph and Imager (ESI; Sheinis et al. 2002)
on the Keck~II 10-m telescope on 2009 March 22.64 UT.  The 0$\farcs$75
slit was used, giving a resolution of $\sim$45~\kms\ over the
range 4000--10,200~\AA.

The two-dimensional (2D) images were processed and the spectra were
extracted using standard routines in IRAF\footnote{IRAF is 
  distributed by the National Optical Astronomy 
  Observatories, which are operated by the Association of Universities 
  for Research in Astronomy, Inc., under cooperative agreement with
  the National Science Foundation.}.  Our own IDL routines were used
to flux calibrate the individual ESI echellette orders and remove telluric
absorption features,
% via comparison to an observation of the standard star G191B2B, 
as well as to combine the individual orders together over
their overlap regions to make the final spectrum.
%The wavelength scale was established using comparison-lamp spectra.
%We found small shifts ($\lesssim0.05$~\AA) of the zeropoints of the
%wavelength scales by cross correlating the night-sky lines in the object
%observations against reference-sky spectra.
The wavelength scale was established using comparison-lamp spectra,
with small shifts ($\lesssim0.05$~\AA) of the zeropoints of the
wavelength scales found by cross correlating the night-sky lines in
the object observations against reference-sky spectra.
The flux level of the first ESI observation was scaled up by 15\%
to match the second before combining them, under the
assumption that the first was affected by clouds or poor guiding.

\section{Results}

\begin{figure}
\plotone{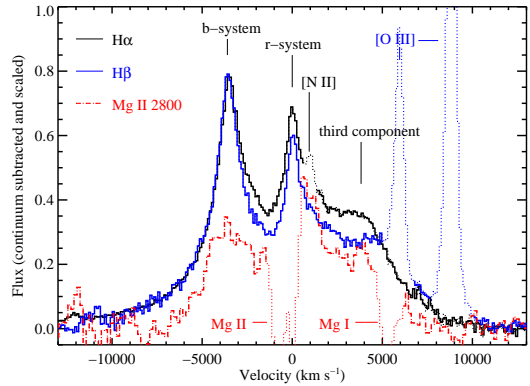}
\caption{Emission-line profiles of \obj.  Shown are H$\alpha$ (black)
  and H$\beta$ (blue) from the Keck/ESI spectrum and \ion{Mg}{2}
  $\lambda$2800 (red) from the Palomar spectrum.  The
  moderate-resolution ESI data have been rebinned for display
  purposes.  A simple spline fit to the continuum beneath each
  emission line has been removed. The zeropoint of the velocity scale
  was set by the narrow [\ion{O}{3}] line associated with the 
  r-system.  Superposed narrow
  emission and absorption lines from other species are shown as dotted
  lines.
}
\label{compfig}
\end{figure}

Both of our spectra reveal that \obj\ has unusual broad-line profiles
(see Fig.~\ref{compfig}), as initially described by \bla. The Balmer 
lines have a
multi-peaked structure, with one peak having the same redshift as the
narrow nebular emission lines ($z=0.3889$) and a second one
blueshifted by $\sim3500$ \kms\ relative to the first.  We follow
\bla\ by referring to these as the ``r-system'' and 
``b-system,'' respectively.  However, as we first noted
\citep{chornockATEL}, the \hal\ line shows a clear third component,
visible as a shoulder or hump present at a velocity of $\sim$4000~\kms\
relative to the r-system, almost symmetrically opposite in velocity
space from the b-system.  The third component is also present in the
\hb\ profile, but it is partially hidden by the superposed [\ion{O}{3}]
$\lambda4959$ emission line.  Although the broad \ion{Mg}{2}
$\lambda2800$ line is contaminated by superposed narrow \ion{Mg}{1}
and \ion{Mg}{2} absorption lines as well as broad \ion{Fe}{2}
emission, it too shows a boxy profile with the ``corners'' of 
the box at the same velocity as the b-system and the third component.

No features due to starlight in the host galaxy are seen in the
spectra, but absorption lines from several common interstellar species
(\ion{Na}{1}, \ion{Ca}{2}, \ion{Mg}{1}, and \ion{Mg}{2}) are present
at a redshift intermediate between the b and r-systems ($z=0.3878$),
as identified by \bla.  Our ESI spectrum, which includes only the
\ion{Na}{1} and \ion{Ca}{2} lines, resolves those absorptions
(full widths at half maximum $\approx$ 100~\kms\ and 170~\kms,
respectively).    Weaker  
(rest-frame equivalent width $\approx$ 0.16~\AA) absorption from
\ion{Na}{1} D2 is also present at $z=0.3890$, close to the redshift of
the r-system, and is likely due to the host galaxy (D1 at this
redshift is blended with D2 of the stronger absorption system). 
The stronger absorption system is therefore blueshifted by 220
\kms\ relative to the narrow emission lines, possibly analogous to the
blueshifted low-ionization ultraviolet (UV) absorber seen in the
prototypical double-peaked emitter, Arp 102B
\citep{halpern96,erac03}.

\begin{figure}
\plotone{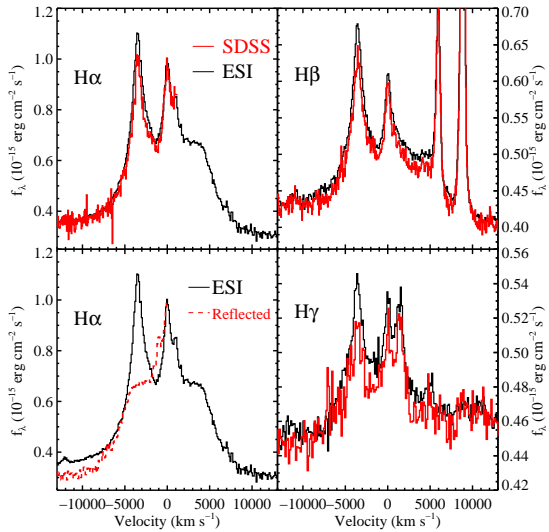}
\caption{Comparison of Balmer-line profiles.  Our Keck/ESI spectrum of \obj\
  is shown in black and the SDSS spectrum analyzed by \bla\ is in
  red. 
% The spectra show very little evolution, except for the peaks
%  near $-3500$~\kms\ (associated with the b-system), visible in
%  \hal, \hb, and H$\gamma$, which are higher in the ESI spectrum than in 
%  SDSS.  
  The lower-left panel compares the blue portion of the
  \hal\ profile in the ESI spectrum (black) to the red half, reflected
  across zero velocity (red-dashed).
}
\label{sdssfig}
\end{figure}

It is of interest to examine the line profiles of \obj\ in more
detail, both to investigate potential line-profile variations and
velocity shifts of emission components.  In Figure~\ref{sdssfig}, we
have plotted a comparison of the Balmer-line profiles in the SDSS and
ESI spectra.  The original SDSS spectrum did not cover the full
\hal\ profile, explaining why \bla\ could not have identified the third
component.  The spectra are very similar, with the biggest
difference being the $\sim$10\% higher peaks associated with the
b-system in the newer spectrum.  This increase in the ratio of the
peak fluxes of the b-system to the r-system is seen consistently in
\hal, \hb, and H$\gamma$. 

The lower-left panel of Figure~\ref{sdssfig} also shows a comparison
of the blue and red halves of the \hal\ profile in our ESI spectrum.
There is perhaps some excess flux at high velocities on the blue side
relative to the red half of the line, but the overall shape of the
wings ($|v|>5000$~\kms) is similar on both sides.  The big difference
between the two halves is the emission peak at $-3500$~\kms\ associated
with the b-system, which is at a very similar velocity (but with the
opposite sign) as the shoulder on the red side.  A potential
explanation is that the 
b-system and the third component represent velocity flows in the
broad-line region that are projected onto our line of sight in
opposite directions (possibly due to rotation or
outflows).  The binary black hole interpretation lacks a natural
explanation for both the existence of the third component and its
presence at an opposite velocity from the b-system.

One explicit prediction of the binary black hole hypothesis made by
\bla\ is that the system might show noticeable velocity shifts due to
the relative orbital motion of the two components of order
100~\kmsy.  Our ESI spectrum was taken 0.95 yr after the SDSS
spectrum, so such a velocity shift should be easily measurable.  We
experimented with several methods to find a robust method of
determining line peaks without being sensitive to assumptions about
the line profile.  We settled on a procedure where we iteratively fit
a quadratic function to each line peak and then fit a quadratic to the
region of $\pm500$~\kms\ around the preliminary peak location.  The
two spectra were fit in an identical manner.  The
differences between the peak wavelengths in the SDSS and ESI
spectra were then converted to velocity shifts and are listed in
Table~\ref{velstab}.  Gaussian fits to the peaks gave similar results
for the shifts.

The two spectra have very consistent wavelength scales, as
demonstrated by the 2.8~\kms\ shift between the [\ion{O}{3}] lines in
the two spectra.  Our ESI spectrum has a high signal-to-noise ratio
(S/N) and velocity resolution, so the errors are dominated by the S/N
of the SDSS spectrum, as demonstrated by the substantially smaller
errors in fitting the bright [\ion{O}{3}] line.  The weighted average
values for the velocity change of the b and r-systems (as measured
from the \hal\ and \hb\ lines) are $-18\pm19$~\kmsy\ and
$16\pm23$~\kmsy, respectively, in the rest frame of the object.  We
can therefore set a 3$\sigma$ limit on any velocity shift of the
underlying emission regions of $\lesssim$70~\kmsy.

\begin{deluxetable}{lc}
\tablecaption{Offsets between SDSS and ESI Spectra}
\tablehead{\colhead{Line} & \colhead{Velocity Shift (\kms)}}
\startdata
[\ion{O}{3}] $\lambda$5007 r-system & -2.8 $\pm$ 2.5 \\
H$\beta$ r-system & 13 $\pm$ 18 \\
H$\alpha$ r-system & 4.8 $\pm$ 27 \\
H$\beta$ b-system & 4.8 $\pm$ 22 \\
H$\alpha$ b-system & -19 $\pm$ 15 \\
\label{velstab}
\enddata
\end{deluxetable}

\citet{wrobel09} observed \obj\ with the VLA and discovered a pair of
radio sources with 0$\farcs$97 separation (named A and B).
Decarli et al. (2009a,b) and \blb\ found $K$-band and optical
sources 
coincident with both the A and B radio components.  One possible
interpretation of these results is that the A and B radio 
sources can be identified with the b and r-systems seen in the
optical spectra, making this source a double quasar with a $\sim$5~kpc
separation. With a fiber size of 3$\arcsec$ in diameter, both systems
would have naturally fallen within the single spectrum obtained by
SDSS. At the time of our ESI observations, we were unaware of the
radio results of \citet{wrobel09} and so did not orient the
spectrographic slit along the vector separating the A+B radio sources
(PA=88$\degr$). Instead, our slit was aligned with the parallactic angle 
(Filippenko 1982; PA=48$\degr$).%, amounting to a relative angular rotation 
%of 40$\degr$.  
We also do not know for certain the exact 
position of the slit on the sky relative to the two components.
%Despite this difficulty, we can still show that the two velocity
%components of the emission lines are spatially unresolved 
%from each other and rule out the double-quasar hypothesis.

\begin{figure}
\plotone{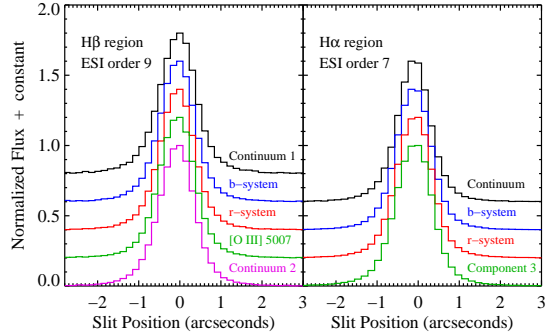}
\caption{Spatial profiles of \obj\ at several different wavelengths.
  The profiles for the order containing
  \hb\ (left panel) represent sums of the counts over wavelength
  intervals corresponding to the peaks of the b and r-systems, the
  [\ion{O}{3}] $\lambda5007$ line, as well as continuum regions on
  either side of the emission complex.  The spatial profiles for the
  order containing \hal\ (right panel) represent the peaks of the
  three components labeled in Figure~\ref{compfig}, as well as a
  continuum region blueward of the emission-line complex. 
}
\label{offsetsfig}
\end{figure}

Whether the slit was centered on the bright optical source
corresponding to the A source or somewhere between the A and B
sources, one would expect an easily measurable spatial offset between
the b and r-systems of 0$\farcs$7 projected along the spatial
direction of the slit if they were associated with the A and B
sources.  The echelle orders of ESI are highly curved 
relative to the CCD columns, so we used the trace of our bright
standard star to rectify each order.  We summed the counts in the
2D CCD frame in several wavelength intervals
corresponding to the three velocity components of the broad emission
lines, as well as the narrow [\ion{O}{3}] $\lambda5007$ line and the
surrounding continuum.  The results for the ESI orders containing
\hal\ and \hb\ are plotted in Figure~\ref{offsetsfig}.

As can be seen from Figure~\ref{offsetsfig}, the spatial profiles are
almost identical at all wavelengths along the slit.  We fit Gaussians
to the spatial profiles and found the centroids to be essentially
coincident.  The maximum and minimum slit positions across the
\hal\ and \hb\ line profiles are equal to
better than 0.1 ESI spatial pixel ($\lesssim0\farcs015$), which is
highly inconsistent with a 0$\farcs$7 projected separation.
Therefore, we reject the double-quasar hypothesis.  \blb\ obtained a
similar constraint on the basis of observations taken with the slit
aligned along a more optimal position angle and also presented
photometric evidence that the B source is associated with an
early-type galaxy.

\section{Discussion}

AGNs with double-peaked broad emission lines have been known for many
years as a relatively rare subclass of AGNs. Systematic surveys of
radio galaxies \citep{eh94} and the SDSS quasar sample
\citep{strateva03} have found numerous members of the class.  The
most plausible current explanation for the displaced red and blue
peaks for most objects in this class involves rotational motion in a
relativistic accretion disk \citep{chen89}, possibly with some
emission component originating in a wind being driven from the surface
of the disk \citep{mc97}. 

Some double-peaked emitters observed in the past have been
proposed to be candidate binary black hole systems
\citep{gaskell83,gaskell96}, much like \obj. 
However, that interpretation has been ruled out in well-studied
systems such as Arp 102B, 3C~390.3, and 3C~332 by the lack of
appropriate long-term velocity evolution of the peaks
\citep{halpern88,eracleous97}.  \bla\ proposed that the observable
velocity evolution would be a test of their black hole 
binary hypothesis.  
%We have determined some upper limits to velocity
%evolution of the peaks in \S3, and \blb\ set similar limits from their
%own data.  While the observed lack of velocity evolution constrains
%the allowed orbital parameters for a potential 
%binary (explored in detail by \blb), we caution that small
%velocity shifts of the line peaks may even be expected in the
%double-peaked emitter scenario.  
While the observed upper limits (\S3; \blb) to the velocity evolution
constrain the allowed orbital parameters for a potential 
binary (explored in detail by \blb), we caution that small
velocity shifts of the line peaks may even be expected in the
double-peaked emitter scenario.  The objects studied by
\citet{eracleous97} did show significant velocity evolution, but in a
manner that was inconsistent with orbital motion.  The origin of these
variations is still not well understood \citep{gezari07}.

The full widths at half maximum and at quarter maximum for
\hal\ in \obj\ (relative to the maximum flux in the central peak) are
8700 and 11,500 \kms, respectively.  These values are almost identical
to the mean values of 8852 and 11,702 \kms\ found by
\citet{strateva03} for their optically selected double-peaked emitter
sample from SDSS, and are significantly larger than those of their
reference sample of normal quasars.

The [\ion{O}{3}] $\lambda$5007 flux in the ESI spectrum is within 2\%
of that in the SDSS spectrum, showing excellent agreement of the
absolute flux scales despite the very different aperture sizes (the
$\lambda$5007 emission region appears to be spatially unresolved;
Fig.~\ref{offsetsfig}).  The total continuum-subtracted \hal\ flux 
over the velocity range $|v| < 10,000$ \kms\ is 1.6 $\times$
10$^{-13}$ ergs cm$^{-2}$ s$^{-1}$, or $L_{{\rm H}{\alpha}} = 8.2 \times 10^{43}$
ergs s$^{-1}$ (assuming a flat $\Lambda$CDM cosmology with $H_0 = 72$ km 
s$^{-1}$ Mpc$^{-1}$;
Dunkley et al. 2009).  This value is higher than any of the
double-peaked emitters from the sample of \citet{strateva03}, but
is exceeded by a few of the radio-selected objects of \citet{eh03},
indicative of the extreme nature of \obj.  In addition, the
low-ionization narrow emission lines have lower rest-frame equivalent
widths ($\sim$0.7~\AA\ for [\ion{O}{1}] $\lambda$6300) than most
double-peaked emitters in the literature \citep{eh03,strateva03}.

\blb\ compared \obj\ with several objects from
the literature, identifying three differences between \obj\ and the
majority of double-peaked emitters.  Two of these differences are the
relative strengths and shapes of the central and blue peaks seen in
\obj\ compared to 
the other objects.  Double-peaked emitters are rather heterogeneous 
as a class \citep{eh94,strateva03}, and we suggest that \obj\ is
an extreme example.  The third difference noted by \blb\
was the lack of change in the intensity of the blue peak over a 
one-year baseline seen in their Kitt Peak spectrum.  However, in our data
(Fig.~\ref{sdssfig}) the blue peak has increased in flux by
$\sim$10\% in both \hal\ and \hb.

\section{Conclusions}

We demonstrated in \S3 that the broad emission components of
\obj\ were spatially coincident, ruling out the double-quasar
hypothesis suggested by \citet{wrobel09} and \citet{decarliIAUC}.
We believe the preponderance of evidence is in favor of interpreting
\obj\ as an unusual double-peaked emitter, not the binary black hole
originally suggested by \bla.  \blb\ have reached similar
conclusions about the nature of this system, but do not yet rule out
the binary hypothesis.
As a phenomenological matter, \obj\ clearly shows multi-peaked
emission-line profiles like other members of the class of
double-peaked emitters.  We 
cannot conclusively disprove the binary black hole hypothesis, but
feel that there is no compelling reason to adopt a binary
interpretation when \obj\ shows characteristics of a previously known
class of objects probably unrelated to binaries.

\obj\ certainly exhibits a few extreme properties for a double-peaked
emitter.  The optical luminosity is very high, as is the ratio of the
blue to red peaks.  \citet{tg09} modeled the Balmer line profiles
in the Keck data presented here and found that a standard disk model
could reproduce the third component and a shelf of emission under the
peak associated with the b-system.  However, their disk models alone
could not reproduce the sharp blue peak, so another component was
required, which they proposed was due to a broad-line region around a
secondary black hole in the accretion disk.  We note that a
significant fraction of the other known double-peaked emitters are
also not well fit in 
detail by simple disk models, either, possibly indicating the presence
of an additional emission component or non-axisymmetric disks
\citep{eh03,strateva03}.

Instead of waiting for several years to further refine the constraints
on the lack of orbital motion in the system as suggested by \blb, we
note that there is potentially a simple test 
that can be done more quickly.  Now that {\it HST}
has been refurbished, it will be possible to obtain UV
spectroscopy.  UV observations of other double-peaked emitters have 
shown that the emission lines from high-ionization material (e.g.,
Ly~$\alpha$ and \ion{C}{4}) frequently lack double-peaked velocity
profiles \citep{halpern96,eracleous04}.  This striking difference
between the profiles of different lines has been attributed to density
and radiative-transfer effects in the multiple line-emitting regions
(e.g., the accretion disk and associated wind).  The absence of
multiple velocity peaks in the UV emission lines would be strong
evidence in favor of \obj\ being a member of the class of double-peaked
emitters; the presence of multiple peaks, however, would still render
the interpretation ambiguous.  UV spectra may also clarify the nature
of the strong blueshifted absorption system.

The high spatial resolution of {\it HST} spectra could also be used 
to obtain separate spectra of both the A and B sources to conclusively
establish the origin of the three components. Likewise,
spatially resolved spectra of the Paschen lines in the
infrared (using adaptive optics from the ground, for instance) should
be fruitful. In either case, we expect such observations to
reveal that either A or B (but not both) would be associated with the
broad emission-line profiles.

\acknowledgments
We would like to acknowledge P. Chang, J. Comerford, M. George,
M. Modjaz, J. Oishi, E. Quataert, and L. Strubbe for lively
conversations during astro-ph coffee, where we first learned of the
\bla\ publication.  We thank J. X. Prochaska and D. Stern for
assistance with the observations.  

Some of the data presented herein were obtained at the W. M. Keck
Observatory, 
which is operated as a scientific partnership among the California
Institute of Technology, the University of California and NASA; it was 
made possible by the generous financial support of the W. M. Keck
Foundation.  A.V.F.'s group at UC Berkeley wishes to acknowledge
financial support from Gary and Cynthia Bengier, the Richard and Rhoda
Goldman Fund, and National Science Foundation grants
AST--0607485 and AST--0908886.

\medskip

{\it Facilities:} \facility{Keck:II (ESI)}, \facility{Palomar:Hale (Double
  Spectrograph)}

\end{document}